\documentclass{article}  
\usepackage{spadre2008}
\usepackage{graphicx}
\frompage{000} \topage{000}                                              
\def\rightmark{Particle Production at RHIC}
\def\leftmark{A Iordanova (STAR Collaboration)}
 
\title{Particle Production at RHIC} 
\authors{
{Aneta Iordanova$^{1,a}$ (for the STAR collaboration) %
}\\[2.812mm]
{\normalsize
\hspace*{-8pt}$^1$ University of Illinois at Chicago, \\ 
Department of Physics, 845 W. Taylor St, M/C 273, 
Chicago, IL, 60607, USA\\[0.2ex] 
}}
 
\abstract{Identified hadron spectra and ratios provide a unique tool
to study the bulk particle production in heavy-ion collisions and
explore the QCD phase diagram.  In these proceedings we present the analysis
of charged pion, kaon and (anti)proton distributions from
$\sqrt{s_{_{NN}}}=$~200 and 62.4 GeV Cu+Cu collisions, collected by
the STAR experiment.  New measurements extend the systematic studies
of bulk properties, addressing the energy and the system size 
 dependence of freeze-out parameters at RHIC.
The available centrality selection of Cu+Cu data bridge the gap
between the smaller d+Au and larger Au+Au systems, allowing a
detailed study of baryon relative to meson and strangeness production as
function of system size.}

\keyword{spectra, bulk production, freeze-out parameters, strangeness} 
\PACS{25.75.-q, 25.75.Ag}
 
\begin{document}
 
\maketitle
\setcounter{page}{1}

\section{Introduction}
Identified low momentum $\pi^{\pm}$, $K^{\pm}$ and $p(\overline{p})$  
particle spectra provide a tool to study the bulk properties 
and to explore the QCD phase 
diagram~\cite{cite:QCD_Diagram}.
The STAR experiment has collected an impressive set of data at
different center-of-mass energies and collision systems.  
In this paper we analyze the Cu+Cu
data at $\sqrt{s_{_{NN}}}=$~200 and 62.4 GeV  over a
broad centrality range, bridging the gap between the smaller
d+Au and larger Au+Au systems and allowing for a detailed study
of hadro-chemistry as a function of system
size.  The measurement of species abundances and their transverse
momentum distributions provide information about the final stages of
the collision evolution at chemical and kinetic freeze-out. 
 
Prior studies of the freeze-out parameters 
in $\sqrt{s_{_{NN}}}=$~200 and 62.4 GeV Au+Au 
collisions~\cite{cite:200spectra,cite:62spectra} within a 
chemical and kinetic equilibrium
model, showed an increasing radial flow with centrality and a similar
chemical freeze-out temperature for both center-of-mass
energies.
Changes in $T_{kin}$ and $\beta$
are found to be consistent with higher energy/pressure in the initial state 
for more central events.
The centrality independence of the extracted chemical freeze-out
temperature was interpreted as medium evolution to the same chemical 
freeze-out, despite differences in the initial conditions. 
Furthermore, the values 
of the chemical freeze-out temperature and the predicted critical temperature
for QCD phase transition, are found to be similar for all
centralities, suggesting that the chemical freeze-out
coincides with hadronization and therefore provides a lower limit
estimate for a temperature of 
prehadronic state~\cite{cite:OlgaPoster}. 
Most measured bulk properties are found to show a smooth systematic 
change with the charged hadron multiplicity and appear to
follow the common systematics with lower-energy collisions data.  
The addition of the
Cu+Cu measurements extends these systematic studies of bulk particle
production at RHIC by addressing not only the energy, but the system
 size and the
inferred energy density dependence of the freeze-out parameters.

\section{Particle identification}

At low-p$_{\rm _{T}}$, charged $\pi^{\pm}$, $K^{\pm}$ and $p(\overline{p})$
are identified using their distinct ionization energy loss ($dE/dx$) 
patterns in the STAR Time Projection Chamber (TPC)~\cite{cite:STAR_TPC}.
The analysis is performed at mid-rapidity ($|y|<0.1$) and 
in six 10\% centrality bins, representing the top 60\% of the
inelastic collision cross-section.  The particle spectra are obtained
from the mean $\langle dE/dx \rangle$ distribution, normalized by the
theoretical expectation for different particle types.  

The normalized distribution is divided into narrow bins of transverse
momentum ($\Delta {\rm p}_{\rm _{T}}=50$~MeV).  For a given momentum and
centrality bin the projections of $dE/dx$ are fit with a four-Gaussian
function, representing the different particle species ($\pi$, $K$,
$p$ and $e$) which can be statistically identified
in this transverse momentum
region ($0.2<{\rm p}_{\rm _{T}}<0.8$~GeV/$c$ for $\pi^{\pm}$,
$k^{\pm}$ and $0.4<{\rm p}_{\rm _{T}}<1.2$~GeV/$c$ for $p(\overline{p})$).
The integral of each Gaussian provides the raw particle yield.
These yields are further corrected for detector
acceptance, tracking inefficiency and background contributions.
The described analysis technique~\cite{cite:200spectra} is consistently
applied to all low-p$_{\rm _{T}}$ particle spectra measurements at both 
$\sqrt{s_{_{NN}}} = 200$ and 62.4~GeV center-of-mass energies and both
Cu+Cu and Au+Au collision systems.

\section{Particle production} 
\subsection{Chemical freeze-out properties}\label{Chem_freeze_out}

In the framework of the statistical model~\cite{cite:StatModel}, ratios of 
particle yields can be
used to provide information about the chemical freeze-out properties of
the system. 
The particle ratios for each centrality bin of Cu+Cu data are fit with
statistical model to derive four parameters: 
the chemical freeze-out temperature (T$_{\rm ch}$), 
the baryon and  strangeness chemical potentials ($\mu_{\rm B}$, $\mu_{\rm S}$)
and the strangeness suppression factor ($\gamma_{\rm S}$).
In this work only $\pi^{\pm}$, $K^{\pm}$,
$p(\bar{p})$ ratios are used \{b\}.

\begin{figure}[h]
\includegraphics[width=0.48\textwidth]{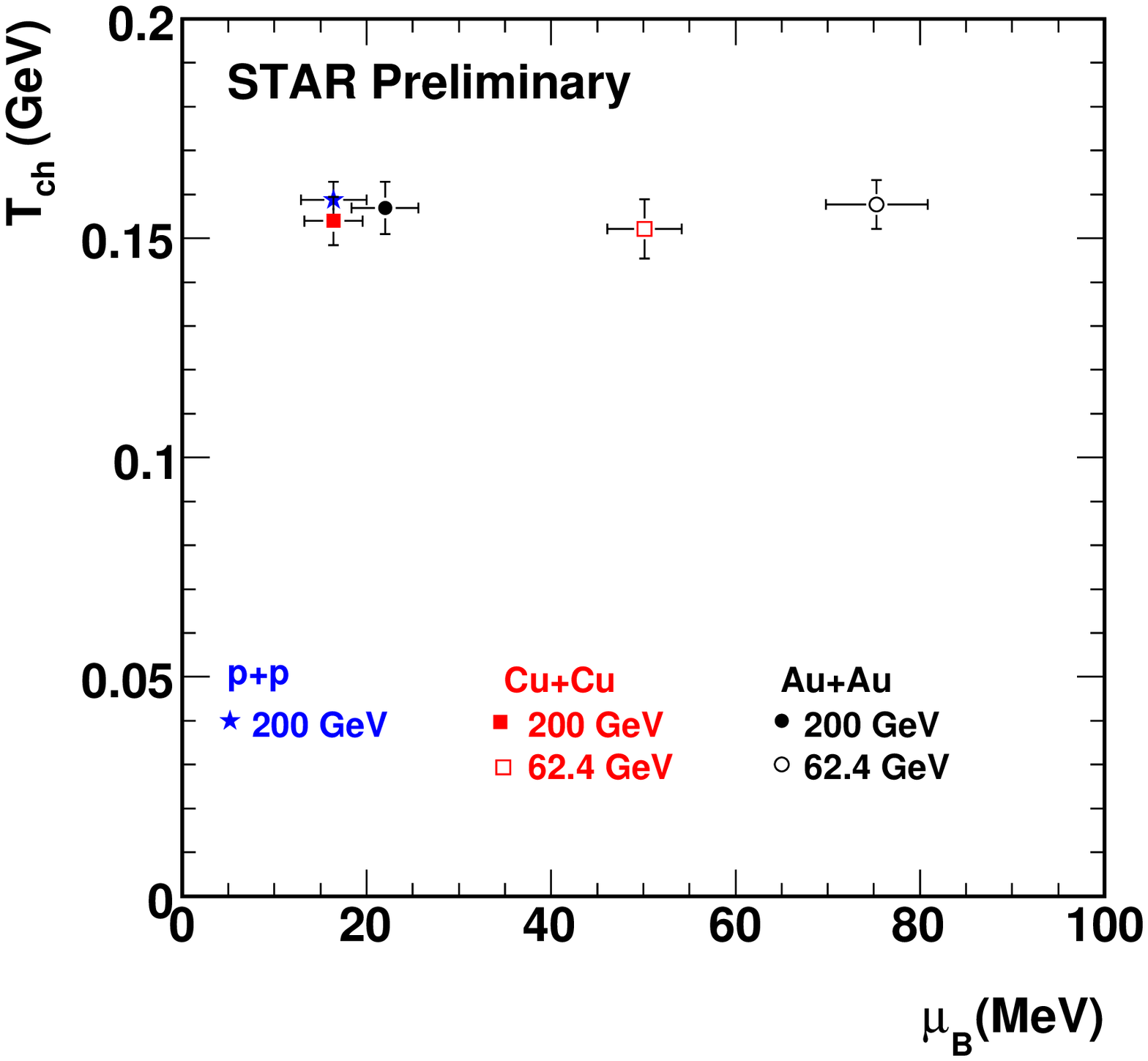}
\includegraphics[width=0.48\textwidth]{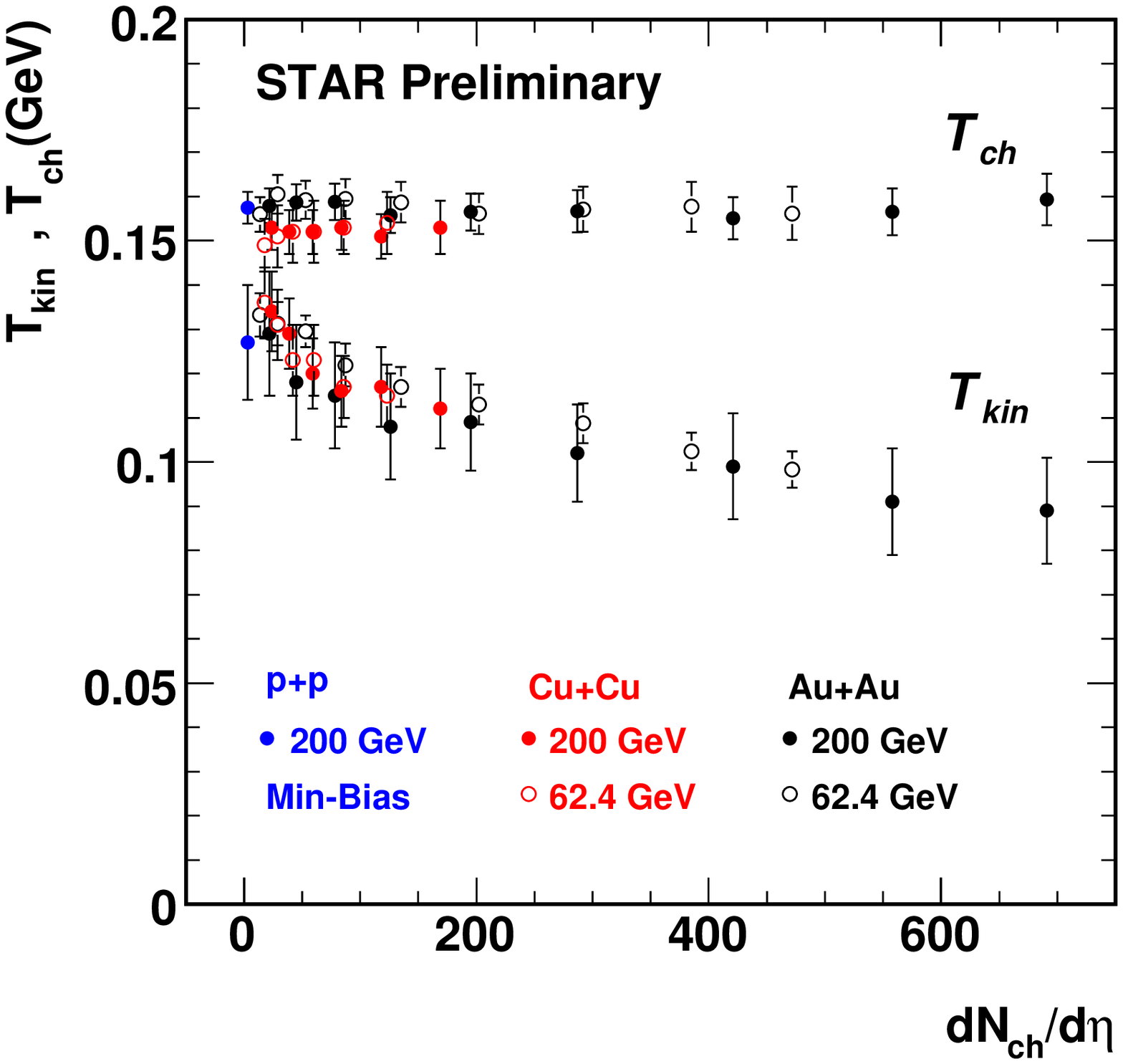}
\caption{\label{fig:ChemicalFreezeOut}
Left panel: Chemical freeze-out temperature, $T_{ch}$,
versus the baryon chemical potential, $\mu_{\rm _{B}}$, for central Au+Au
(0-5\%, circles) and Cu+Cu collisions (0-10\%, squares).  Minimum-bias p+p
data at 200~GeV are also shown (star).  
Right panel: $T_{ch}$ ($T_{kin}$)
versus charged hadron multiplicity at $\sqrt{s_{_{NN}}}$~62.4 (open symbols)
and 200~GeV (closed symbols) for Cu+Cu and Au+Au  collisions.
For comparison, results for minimum-bias p+p collisions at 200 GeV are
also shown.}
\end{figure}

The chemical freeze-out temperature 
as a function of baryon-chemical potential 
for different systems and colliding energies is shown in the left panel of
Fig.~\ref{fig:ChemicalFreezeOut}. 

The value of T$_{\rm ch}$ appears to be universal. For all systems
and center-of-mass energies T$_{\rm ch}$ is constant as a function
of $\mu_{\rm B}$ as well as the charged particle mid-rapidity multiplicity
($dN_{ch}/d\eta$), right panel of Fig.~\ref{fig:ChemicalFreezeOut}.
The constant value of T$_{\rm ch}$ for collisions with different
initial conditions (energy and net-baryon densities) points to 
a common hadronization temperature across the systems studied.
The value of the baryon chemical
potential reflects the decrease in baryon density from
$\sqrt{s_{NN}}=$~62.4 to 200~GeV.  At the same center-of-mass energy
$\mu_{\rm B}$ is higher for the larger system. 

The strangeness suppression factor $\gamma_{\rm S}$ in Cu+Cu is consistent 
with the results for the Au+Au data~\cite{cite:200spectra} within the 
systematic errors on the fit parameters.
This parameter
shows a similar dependence with $dN_{ch}/d\eta$, as observed in the
Au+Au system.  The value of $\gamma_{\rm S}$ approaching unity for
the central collisions implies that the produced strange quark is in
approximate equilibrium with $u$ and $d$ quarks.

\subsection{Kinetic freeze-out properties}\label{Kin_freeze_out}

To characterize the final freeze-out state
we fit all particle spectra
within a given centrality bin by the Blast-wave 
model~\cite{cite:BlastWaveModel}.
The model assumes a radially expanding thermal source.  The
hydro-motivated fits provide information about the radial flow 
velocity ($\beta$) and the kinetic freeze-out temperature
(T$_{\rm kin}$) at final freeze-out.  The effects from resonance
contributions to the pion spectral shape are reduced by excluding
the very low-p$_{\rm _{T}}$ pion data points ($<$~0.5~GeV/${c}$) \{c\}. 

The particle spectra are well described by a common set of freeze-out
parameters for all colliding energies.
When measured in collisions with 
similar $dN_{ch}/d\eta$,  T$_{\rm kin}$ and $\beta$ exhibit similar
centrality dependences in both Cu+Cu and Au+Au
collisions, evolving smoothly from the lowest (p+p) to the highest 
(central Au+Au) available multiplicities. T$_{\rm kin}$ decreases
with centrality and thus indicating that the freeze-out occurs at a
lower temperature in more central collisions (see the right panel
of Fig.~\ref{fig:ChemicalFreezeOut}).  
The particle mean-p$_{\rm _{T}}$ 
increase with $dN_{ch}/d\eta$, which is consistent with an increase 
in radial flow  $\beta$ with centrality~\cite{cite:SQM07proc}. 

In conjunction with Color Glass Condensate model~\cite{cite:CGC} 
the number of produced charged particles can be connected with 
the initial gluon density of the colliding
system, leading to the interpretation that the bulk freeze-out
properties are most probably determined at the initial stages of the
collision and are driven by the initial energy density.

\subsection{Baryon and meson production}

\begin{figure}[h]
\begin{minipage}{15pc}
\includegraphics[width=15pc]{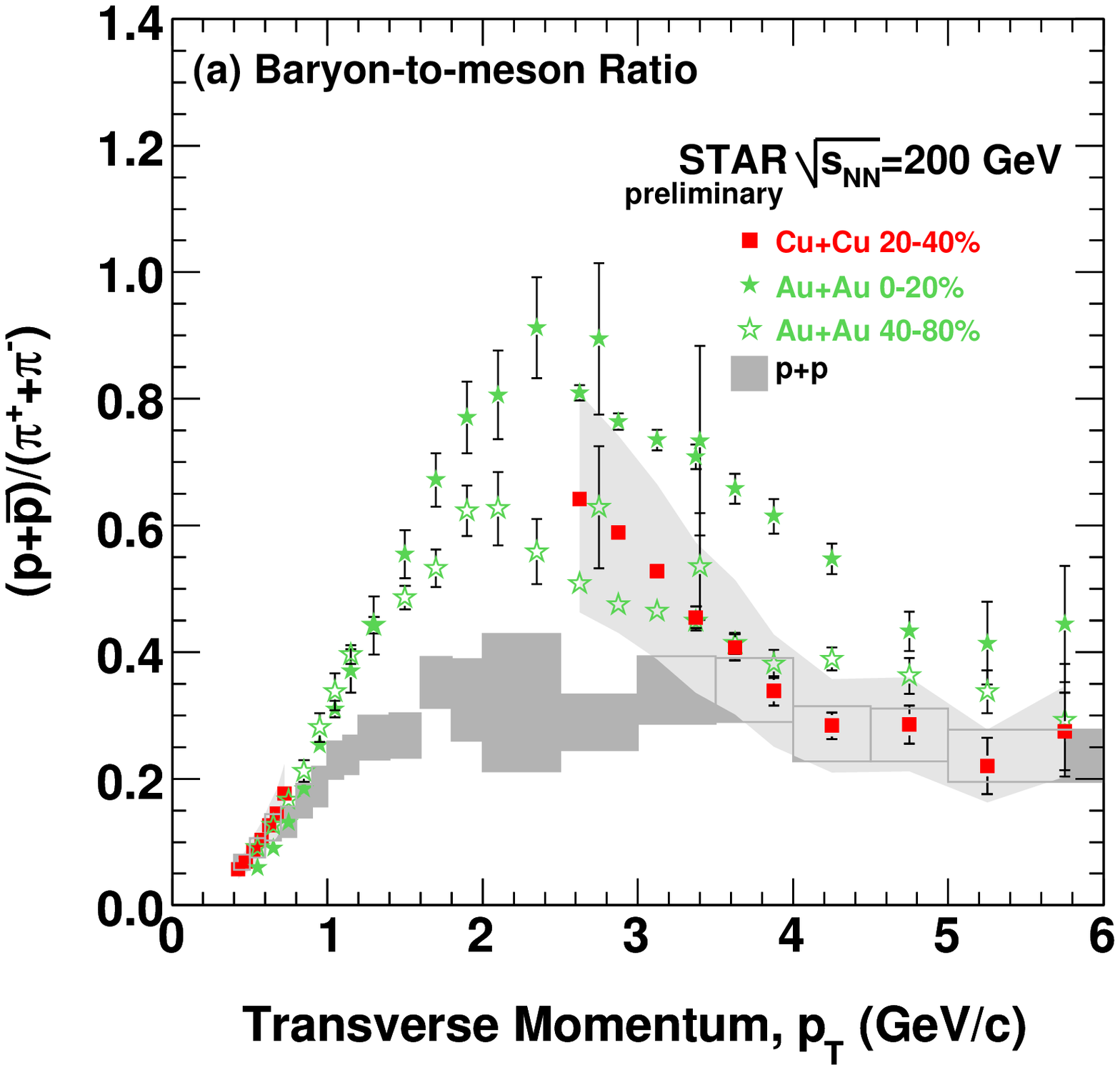}
\end{minipage}
\begin{minipage}{15pc}
\includegraphics[width=15pc]{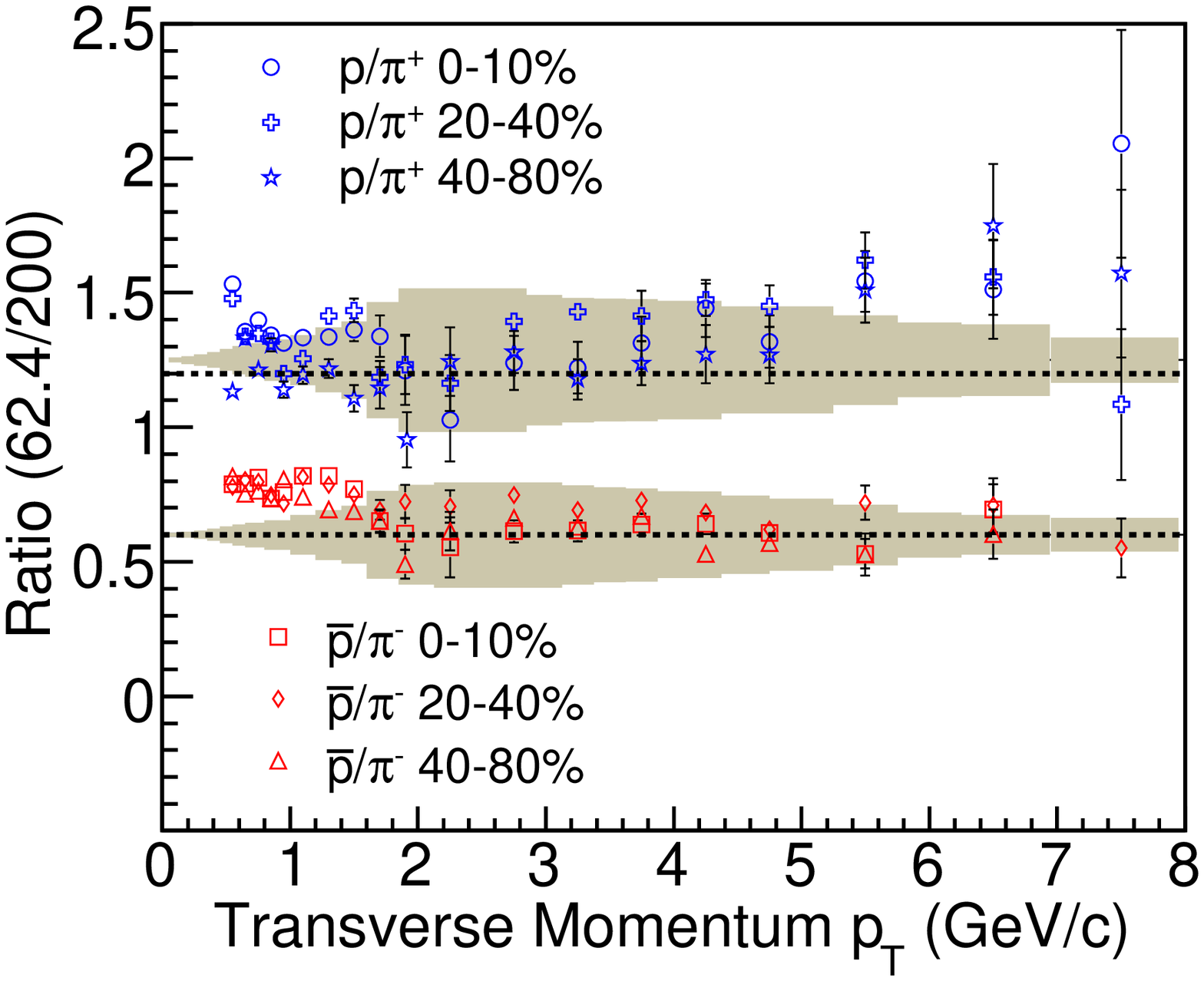}
\end{minipage}
\caption{\label{fig:Baryon_to_meson}
 Left panel: The baryon to meson ratio
(p+$\overline{\rm p}$/$\pi^{+}+\pi^{-}$) as a function of transverse
momentum p$_{\rm _{T}}$ for Cu+Cu (squares) and Au+Au (stars) for
different centralities. The shaded area represents the ratio in
p+p collisions.  
Data are from \cite{cite:200spectra,cite:PRL97,cite:WWND07_RH,cite:PLB_ppdAu_low,cite:PLB_ppdAu_high}.
Right panel: The p/$\pi^{+}$ and
$\overline{\rm p}$/$\pi^{-}$ ratios in 62.4 compared to 200~GeV Au+Au
collisions as a function of transverse momentum.  Different symbols
represent centrality bins.  Data are from \cite{cite:PLB655}.}
\end{figure}

The ratio of
baryons (inclusive $p+\overline{p}$) to mesons ( $\pi^{+} + \pi^{-}$) 
as a function of 
p$_{\rm _{T}}$ for p+p, Cu+Cu and Au+Au systems at $\sqrt{s_{_{NN}}}=$~200~GeV
is shown in Fig.~\ref{fig:Baryon_to_meson} (left panel).
At all momenta the number of baryons is less than the number of
produced mesons. There is no centrality dependence for low transverse
momenta and the relative production in A+A is very similar to that in
elementary p+p collisions.  At intermediate-p$_{\rm _{T}}$ the baryon
to meson ratio is enhanced relative to the p+p system. A strong
centrality dependence is observed, with the ratio reaching a maximum
for central collisions at p$_{\rm _{T}} \sim $2~GeV/$c$.  For
p$_{\rm _{T}}>5$~GeV/$c$ in Cu+Cu (p$_{\rm _{T}}>$~7~GeV/$c$ in Au+Au) the
baryon to meson production is found to be the same as p+p and independent 
of centrality.
Ratios formed from particles containing strange quarks
($\Lambda$/K$^{0}_{S}$)~\cite{cite:STAR_Lambda_K0s} show similar
features in the baryon to meson production as a function of
p$_{\rm _{T}}$ and centrality. 

The baryon to meson production at different energies in Au+Au data is
studied in the right panel of Fig.~\ref{fig:Baryon_to_meson}. The
ratio of protons to pions (and $\overline{p} / \pi^{-}$) at 62.4 GeV 
is compared to
that in 200~GeV.  The baryon to meson production at a given centrality
in 62.4~GeV has the same transverse momentum dependence as that in 200~GeV.

\subsection{Strangeness production}

The ratio of charged kaon and pion yields in Cu+Cu and Au+Au allows
us to gain some insight about the strangeness production as a function of the 
system size.  The centrality dependence  of $K/\pi$ ratio in Cu+Cu collisions
shown as function of 
$dN_{ch}/d\eta$, \{d\} follows the same trend as previously
found in Au+Au data at the top RHIC 
energy~\cite{cite:200spectra,cite:SQM07proc}. 
There is no strong evidence for additional strangeness enhancement
of the kaon yield relative to the pion reference in the smaller system,
as observed at AGS and SPS 
energies~\cite{cite:SPS_NA49QM02,cite:SPS_NA49PRL},
despite the observed increase in the integrated particle spectra
yields with respect to p+p data for a given value of
$N_{part}$~\cite{cite:ATimmins_SQM07}.

\section{Summary}
The STAR collaboration has presented measurements of identified charged
hadron spectra in Cu+Cu collisions for two center-of-mass energies, 200
and 62.4~GeV.  These new results of $\pi^{\pm}$, $K^{\pm}$, $p(\bar{p})$
have further enriched the variety of low-$p_{T}$ spectra measurements at RHIC.
The data have been studied within the frameworks of statistical and 
Blast-wave model
in order to characterize the properties of the final hadronic
state of the colliding system 
and explore the freeze-out systematics
as a function of system size, collision
energy, centrality and the inferred energy density.

This multi-dimensional systematic study reveals remarkable similarities
between the systems studied. In both Cu+Cu and Au+Au collisions,
mid-rapidity  baryon production is enhanced compared to that of mesons
at intermediate $p_{T}$ indicating the coalescence process for hadronization.
There is no strong evidence for additional strangeness enhancement in 
the smaller Cu+Cu system in comparison to Au+Au. 
When measured in collisions with  similar $dN_{ch}/d\eta$,  
T$_{\rm kin}$ and $\beta$ exhibit similar
centrality dependences in both Cu+Cu and Au+Au collisions. 

The obtained freeze-out parameters are found to be intrinsically
related for all collision systems and center-of-mass energies.  A
smooth evolution with $dN_{ch}/d\eta$ and similar properties at the same
number of produced charged hadrons are observed.  
The assumption that the number of produced charged particles is 
representative of the initial gluon density of the colliding 
system~\cite{cite:CGC}  can be used to interpret that the bulk 
freeze-out properties are most
probably determined at the initial stages of the collision and are
driven by the initial energy density.

 
\section*{Notes}
\begin{notes}
\item[a]
E-mail: aiorda1@uic.edu

\item[b]
For results using multi-hadron fits see~\cite{cite:Jun_Takahashi}.

\item[c]
Studies show that inclusion of resonance decays (at RHIC energies) 
do not affect significantly the fit results.

\item[d] 
When compared at the same center-of-mass energy, the ``$k/p$'' ratios are 
the same in both centrality representations, $dN_{ch}/d\eta$ and $N_{part}$.
\end{notes}

\vfill\eject
\end{document}